\documentclass[12pt]{iopart}
\usepackage{harvard}
\usepackage{graphicx}
\newcommand{\be}{\begin{equation}}
\newcommand{\ee}{\end{equation}}

\newcommand{\Mc}{{\cal M}}
\newcommand{\Ms}{M_{\odot}}

\begin{document}

\title[Sky Localisation of Gravitational Waves]{Degeneracies in Sky Localisation Determination from a Spinning Coalescing Binary through Gravitational Wave Observations: a Markov-Chain Monte-Carlo Analysis for two Detectors.}

\author{V. Raymond$^{1}$, M.V.\ van der Sluys$^{1}$, I.\ Mandel$^{1}$, V.\ Kalogera$^{1}$, C.\ R\"{o}ver$^{2,3}$, N.\ Christensen$^{4}$}

\address{$^1$ Dept. of Physics \& Astronomy, Northwestern University, 2131 Tech Drive,  Evanston IL, 60208,  USA} 
\address{$^2$ Department of Statistics, University of Auckland, Private Bag 92019, Auckland 1142, New Zealand}
\address{$^3$ Max-Planck-Institut f\"{u}r Gravitationsphysik, Callinstra\ss{}e 38, 30167 Hannover, Germany}
\address{$^4$ Physics \& Astronomy Dept., Carleton College, One North College Street, Northfield MN 55057, USA}

\ead{vivien@u.northwestern.edu}

\begin{abstract}
Gravitational-wave signals from inspirals of binary compact objects (black holes and neutron stars) are primary targets of the ongoing searches by ground-based gravitational-wave interferometers (LIGO, Virgo, and GEO-600). 
We present parameter-estimation simulations for inspirals of black-hole--neutron-star 
binaries using Markov-chain Monte-Carlo methods.
As a specific example of the power of these methods, we consider
source localisation in the sky and analyse the degeneracy in it when data from only two detectors are used. We focus 
on the effect that the black-hole spin has on the localisation estimation.
We also report on a comparative Markov-chain Monte-Carlo analysis with two different waveform
families, at 1.5 and 3.5 post-Newtonian order. 
\end{abstract}

\maketitle

\section{Introduction}
\label{sec:intro}

Binary systems with compact objects --- neutron stars (NSs) and black holes (BHs) --- in the mass 
range $\sim 1\,\Ms - 100\,\Ms$ are among the most likely sources of gravitational waves (GWs) 
for ground-based laser interferometers currently in operation 
\cite{2002gr.qc.....4090C}: LIGO \cite{1999PhT....52j..44B}, Virgo \cite{2004CQGra..21..385A} and GEO-600 \cite{2004CQGra..21S.417W}.
Merger-rate estimates are quite uncertain and for BH-NS binaries current detection-rate 
estimates range
from 0.0003 to 0.1\,yr$^{-1}$ 
for first-generation instruments \citeaffixed{2008ApJ...672..479O}{\emph{e.g.}}. 
Upgrades to Enhanced LIGO/Virgo (2008--2009) and Advanced LIGO/Virgo (2011--2014)
are expected to increase detection rates by factors of about $\sim 10$ and $\sim 10^3$, respectively. 

The measurement of source properties holds major promise for 
improving our astrophysical understanding and requires reliable methods for parameter estimation. 
This is a challenging problem, however, because of the large 
number of parameters ($9$ for circular non-spinning binaries, and more for spinning systems) and the significant amount of structure in the parameter space. In the case of low-mass-ratio binaries (\emph{e.g.}\ BH-NS),
these issues are amplified 
for significant spin magnitudes and large misalignments between the BH spin and the orbital angular momentum
\cite{1994PhRvD..49.6274A,2003PhRvD..67d2003G,2003PhRvD..67j4025B}. However, the presence of  
spins improves parameter estimation through the signal modulations, although still  
presenting us with a considerable computational challenge. This was highlighted  
in the context of LISA observations \citeaffixed{2004PhRvD..70d2001V,2006PhRvD..74l2001L}{see}
and in our first study devoted to ground-based observations \cite{2008ApJ...688L..61V}.

In this paper we examine the potential for parameter estimation of  
spinning binary inspirals with ground-based interferometers. 
\citename{2006CQGra..23.4895R} \citeyear{2006CQGra..23.4895R,2007PhRvD..75f2004R}
explored parameter estimation for non-spinning binaries, which requires 9 parameters. We focus on BH-NS binaries, 
which can exhibit significant coupling between the orbital angular momentum and the BH spin, mainly because of 
the high mass ratio \cite{1994PhRvD..49.6274A}, while at the same time we are justified to ignore the NS spin, 
leading to a 12-dimensional parameter space.
We apply a newly developed Markov-chain Monte-Carlo (MCMC) algorithm \cite{2008CQGra..25r4011V} to spinning inspiral signals injected into synthetic noise and we derive posterior  
probability-density functions (PDFs) of all twelve signal parameters. 
In our previous study \cite{2008ApJ...688L..61V}, we showed the accuracy obtained in sky-position determination 
using data from a two-detector network, where a degeneracy in the sky position exists, and from a three-detector network, where the degeneracy is broken.
Following this work, we here analyze in further detail the degeneracy which is present when data from only two detectors are used.
In section~\ref{sec:skyring} we show that the degeneracy in the sky 
position is reduced but not lifted when a significant spin is present ($a_\mathrm{spin}\ge0.5$), and that a 
sufficient angle between spin and orbital angular momentum can break such a degeneracy 
($\theta_{SL}\ge55^{\circ}$). In this study, we demonstrate that these degeneracies are due to 
a high degree of similarity between signals from sources with significantly different parameter 
sets, while we made sure the observed effects are real and
not in fact artifacts due to potential errors in our analysis
methodology.

In section~\,\ref{sec:comparison}, we also demonstrate that the inclusion of 
higher post-Newtonian orders in the waveform can improve the accuracy of intrinsic-parameter estimation.  Meanwhile, we find 
that the additional structure in the parameter space of higher-order waveforms lowers the sampling efficiency of the MCMC and requires improvements to the sampling scheme.

\section{Signal and observables}
\label{sec:GW}

In this paper we analyze the signal produced during the inspiral phase of two  
compact objects of masses $M_{1,2}$ in circular orbit. We  
focus on a fiducial BH-NS binary system with $M_1 = 10\,\Ms$ and $M_2 = 1.4\,\Ms$, and we ignore the NS spin. The BH spin $\mathbf{S}$
couples to the orbital angular momentum $\mathbf{L}$, leading to amplitude and phase modulations  
of the observed gravitational radiation due to the precession of the orbital plane. 
Here we model GWs by post-Newtonian (pN) waveforms, either at ${1.5}$-pN order in phase and Newtonian 
amplitude or at ${3.5}$-pN order in phase and Newtonian amplitude. For the latter waveform 
we use the implementation from the LSC Algorithm Library (LAL) \cite{LAL}. 
In our analysis we model the noise in each detector as a zero-mean Gaussian, stationary random process, 
with one-sided noise spectral density $S_a(f)$, 
at the initial-LIGO design sensitivity 
\cite{2003SPIE.4856..282F}.

\subsection{Waveform template at the ${1.5}$-pN order }
\label{sec:1.5pn-template}

We adopt the \emph{simple-precession} limit \citeaffixed{1994PhRvD..49.6274A}{Eqs.~51, 52, 59 \& 63 in},
appropriate for the single-spin system considered here. For simplicity (to speed up the  
waveform calculation), we also ignore the \emph{Thomas precession} \cite{1994PhRvD..49.6274A}.
In this approximation, the orbital angular momentum $\mathbf{L}$ and
spin $\mathbf{S}$ precess with the \emph{same} angular frequency 
around a fixed direction $\hat{\mathbf{J}}_0 \approx \hat{\mathbf{J}}$, where $\mathbf{J} = \mathbf{L} + \mathbf{S}$.
During the inspiral phase the spin misalignment
 $\theta_{SL} \equiv \mathrm{arccos}({\bf \hat S} \cdot {\bf \hat L})$ and $S = |\mathbf{S}|$ 
are constant. These approximate waveforms retain (at the leading  
order) all the prominent qualitative features introduced by the spins, while allowing us to rapidly 
compute the waveforms analytically. 
While this approach is justified for the exploration of GW astronomy and the development of  
parameter-estimation algorithms, more accurate waveforms 
\citeaffixed{1995PhRvD..52..821K,1996PhRvD..54.4813W,2006PhRvD..74j4033F,2006PhRvD..74j4034B}{\emph{e.g.}}
will be necessary for the analysis of real signals (see section~\,\ref{sec:comparison}). 

A circular binary inspiral with one spinning compact object is described by a 12-dimensional parameter  
vector $\vec{\lambda}$. With respect to a fixed geocentric coordinate system our choice of  
independent parameters is:
\be
\vec{\lambda} = \{\Mc,\eta,\mathrm{R.A.},\sin\mathrm{Dec},\sin\theta_{J_0},\phi_{J_0},\log{d_\mathrm{L}},a_\mathrm{spin},\cos\theta_{SL}, \phi_\mathrm{c},\alpha_\mathrm{c}, t_\mathrm{c}\},
\label{e:lambda}
\ee
where $\Mc = \frac{(M_1 M_2)^{3/5}}{(M_1 + M_2)^{1/5}}$ and $\eta = \frac{M_1 M_2}{(M_1 + M_2)^2}$ are  
the chirp mass and symmetric mass ratio, respectively; R.A. (right ascension)  
and Dec (declination) identify the source position in the sky;  
the angles $\theta_{J_0} \in \left[-\frac{\pi}{2},\frac{\pi}{2} \right]$  
and $\phi_{J_0}\in \left[0, 2\pi \right[$ identify the unit vector $\hat{\mathbf{J}}_0$; $d_\mathrm{L}$ is the  
luminosity distance to the source and $0 \le a_\mathrm{spin} \equiv S/M_1^2 \le 1$ is the  
dimensionless spin magnitude; $\phi_\mathrm{c}$ and $\alpha_\mathrm{c}$ are integration constants that specify  
the GW phase and the location of $\mathbf{S}$ on the precession cone, respectively, at 
the time of coalescence $t_\mathrm{c}$, defined with respect to the center of the Earth. 

Given a network comprising $n_\mathrm{det}$ detectors, the data collected at the $a-$th  
instrument ($a = 1,\dots, n_\mathrm{det}$) is given by $x_a(t) = n_a(t) + h_a(t;\vec{\lambda})$,  
where $h_a(t;\vec{\lambda}) = F_{a,+}(t)\,h_{a,+}(t;\vec{\lambda}) + F_{a,\times}(t)\,h_{a,\times}(t;\vec{\lambda})$  
is the GW strain at the detector \citeaffixed{1994PhRvD..49.6274A}{see Eqs.\,2--5 in}
and $n_a(t)$ is the detector noise. The astrophysical signal  
is given by the linear combination of the two independent polarisations $h_{a,+}(t;\vec{\lambda})$  
and $h_{a,\times}(t;\vec{\lambda})$ weighted by the \emph{time-dependent} antenna beam  
patterns $F_{a,+}(t)$ and $F_{a,\times}(t)$.

\subsection{Waveform template in the ${3.5}$-pN order}

\begin{figure}
  \includegraphics[angle = 270, width=\linewidth]{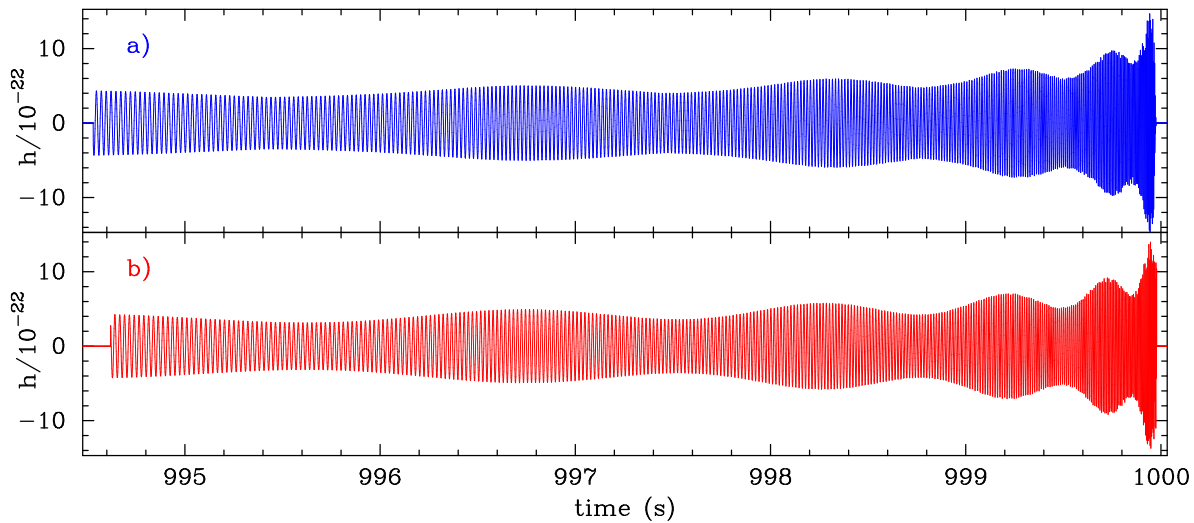}
  \caption{
    {\bf (a)} Part of the ${1.5}$-pN time-domain waveform from a source with $a_\mathrm{spin}=0.5$ and  
    $\theta_{SL}=20^\circ$.   
    {\bf (b)} The ${3.5}$-pN waveform from a source with the same parameters. 
    The waveforms start at 40\,Hz and are aligned at the coalescence time.
    \label{fig:wave}
  }
\end{figure}

Although the ${1.5}$-pN, simple-precession waveform is useful to investigate 
the principal effects of spin on parameter estimation, a more realistic waveform is needed to analyse detected signals.
The waveform we use for this is ${3.5}$-pN in phase and Newtonian amplitude. 
The implementation in the LSC Algorithm Library closely follows the first section of \citename{2003PhRvD..67j4025B} \citeyear{2003PhRvD..67j4025B}.
For comparison purposes we converted the usual set of parameters used in the LAL software 
to the parameters in Eq.\,\ref{e:lambda}. In doing so, we fix 3 of the 15 parameters of the LAL parameter set, 
setting the spin of the second member of the binary to be 0.
The waveform is generated using \texttt{LALGenerateInspiral()} from the injection package \cite{LAL}. 
An example of $h_a(t)$ for $a_\mathrm{spin}=0.5$ and $\theta_{SL}=20^\circ$ for both waveforms 
is shown in figure~\,\ref{fig:wave}. 

\section{Parameter estimation: Methods}
\label{sec:methods}

The goal of our analysis is to determine the \emph{posterior} probability density function (PDF) of the unknown parameter  
vector $\vec{\lambda}$ in Eq.\,\ref{e:lambda}, given the data sets $x_a$ collected by a  
network of $n_\mathrm{det}$ detectors and the \emph{prior} $p(\vec{\lambda})$ on the  
parameters.  We use wide, flat priors (see \citeasnoun{2008CQGra..25r4011V} for details).
Bayes' theorem provides a rigorous mathematical rule to assign such a probability: 
\be
p(\vec{\lambda}|x_a ) = \frac{p(\vec{\lambda}) \, {\cal L} (x_a|\vec{\lambda})}{p(x_a)}\,;
\label{e:jointPDF}
\ee
in the previous equation
\be
{\cal L}(x_a|\vec{\lambda}) \propto 
\exp\left(
<x_a|h_a(\vec{\lambda})>-\frac{1}{2}<h_a(\vec{\lambda})|h_a(\vec{\lambda})>
\right)
\label{e:La}
\ee
is the \emph{likelihood function} of the data given the model, 
which measures the fit of the data to the model, and
$p(x_a)$ is the  \emph{marginal likelihood} or \emph{evidence}.  We use the notation
\be
<x|y>=4Re\left( \int_{f_{\rm low}}^{f_{\rm high}}\frac{\tilde{x}(f)\tilde{y}^{*}(f)}{S_a(f)}\,\mathrm{d}f \right)
\label{e:prod}
\ee
to denote the \emph{overlap} of signals $x$ and $y$, where $\tilde x(f)$ is the Fourier transform of $x(t)$.  
For future reference, we also define match between two waveforms corresponding to different parameter values as the overlap between the normalized waveforms:
\be
M(h(\vec{\lambda_1}),h(\vec{\lambda_2}))=\frac{<h(\vec{\lambda_1})|h(\vec{\lambda_2})>}{\sqrt{<h(\vec{\lambda_1})|h(\vec{\lambda_1})><h(\vec{\lambda_2})|h(\vec{\lambda_2})>}} 
\label{e:M}
\ee
Note that with the definitions employed here, the ``true'' likelihood computed at the parameters of the injected waveform ${\cal L_{\rm true}}(x|\vec{\lambda}_{\rm true})$ is a random variable that depends on a particular realization of the noise $n_a$ in the data $x_a=h(\vec{\lambda}_{\rm true})+n_a$.  We define the signal to noise ratio (SNR) to be the square root of twice the expectation value of $\log {\cal L_{\rm true}}$:
\be
SNR= \sqrt{<h(\vec{\lambda}_{\rm true})|h(\vec{\lambda}_{\rm true})>}
\label{e:SNR}
\ee

For multi-detector observations involving a network of detectors with  
uncorrelated noise 
--- this is the case of this paper, where we use up to two non-colocated detectors ---
we have ${\cal L}(\vec{x}|\vec{\lambda}) = \prod_{a=1}^{n_\mathrm{det}}\, {\cal L}(x_a|\vec{\lambda})\,$, for $\vec{x} \equiv \{x_a: a = 1,\dots,n_\mathrm{det}\}$ and:
\be
p(\vec{\lambda}|\vec{x}) = \frac{p(\vec{\lambda})\, {\cal L}(\vec{x}|\vec{\lambda})}
{p(\vec{x})}
\ee
The numerical computation of the joint and \emph{marginalised} PDFs involves the  
evaluation of integrals over a large number of dimensions. Markov-chain Monte-Carlo  
(MCMC) methods \citeaffixed[and references therein]{gilks_etal_1996,gelman_etal_1997}{\emph{e.g.}}
have proved to be especially effective in tackling these numerical problems.
We developed an adaptive \citeaffixed{figueiredo_jain_2002,atchade_rosenthal_2005}{see}
MCMC algorithm to explore the parameter  
space efficiently while requiring the least amount of tuning for the specific signal at  
hand; the code is an extension of the one developed by some of the authors to explore  
MCMC methods for non-spinning binaries \cite{2006CQGra..23.4895R,2007PhRvD..75f2004R}
and takes advantage of techniques explored by some of us in the context of LISA data 
analysis \cite{2007CQGra..24..541S}.  
A summary of the methods used in our MCMC code was published in \citeasnoun{2008CQGra..25r4011V};
more technical details will be provided elsewhere.

\section{Parameter estimation: Results}
\label{sec:results}

\subsection{MCMC runs}
\label{sec:runs}

Here we present results obtained by injecting a signal into simulated interferometer noise 
and computing the posterior PDFs with MCMC techniques, for a fiducial  
source consisting of a $10\,M_\odot$ spinning BH and a $1.4\,M_\odot$ 
non-spinning NS in a binary system at a distance of about 16\,Mpc (see sections~\,\ref{sec:skyring} and 
\ref{sec:comparison} for parameters values). We consider a number of cases for which we change the BH spin parameters. 
We run the analysis using the simulated data from (i) the 4-km LIGO detector at Hanford (H1) alone, (ii) the LIGO Hanford (H1) and a second detector with the initial-LIGO noise curve located and oriented in the same way as the Virgo detector near Pisa, which we denote by (P).

The MCMC analysis that we carry out on each data set consists of 5 to 25 independent chains,
each with a length of several millions iterations.
The chains are sampled after a \emph{burn-in} period \citeaffixed{gilks_etal_1996}{see \emph{e.g.}}
that is determined
automatically as follows: we determine the absolute maximum likelihood $\log({\cal L}_\mathrm{max})$, 
defined as the highest likelihood $\log[{\cal L}(\vec{x}|\vec{\lambda})]$ obtained over the ensemble of parameters 
$\vec{\lambda}$ for which the overlap has been computed, for any of the individual chains. Then for
each chain we include all the iterations {\it after} the chain reaches a likelihood value of 
$\log({\cal L}_\mathrm{max})-2$ for the first time.
All our Markov chains start at offset ({\it i.e.}, non-true) parameter values, unless specified otherwise. The starting values for 
$\Mc$ and $t_\mathrm{c}$ 
are drawn from a Gaussian distribution 
centred on the true parameter value, with a standard deviation of $0.1\,M_\odot$ and 
30\,ms respectively.  The other ten parameters are drawn uniformly from the allowed ranges.
Our MCMC code needs to run for typically a few days to one week in order to show the first results and 10--14 days
to accumulate a sufficient number of iterations for good statistics, each chain using
a single 2.8\,GHz CPU.

\subsection{A Ring in the Sky}
\label{sec:skyring}

For the study in this section, we use the ${1.5}$-pN waveform with the \textit{simple-precession}
prescription only (see section~\,\ref{sec:1.5pn-template}).  In order to further speed up the MCMC runs,
we fixed the mass and spin parameters to the true parameter values, and performed test calculations
to verify that this does not affect our conclusions.

As reference MCMC runs (experiment 1, see table \ref{table:exp}), we injected signals 
into two simulated detectors (H1 and P).  We made three different injections, which differed in the spin magnitude values  ($a_\mathrm{spin} = 0.0, 0.5$) and the spin-misalignment values ($\theta_{SL}=20^\circ, 55^\circ$) for $a_\mathrm{spin} = 0.5$ case; the remaining parameters were kept constant across all three injections: $\Mc=3.0 M_\odot$, $\eta=0.11$ ($M_1=10.0 M_\odot$, $M_2=1.4 M_\odot$), 
$d_\mathrm{L}$ = 16\,Mpc, R.A.\ = 14.3\,h, Dec = 11.5$^\circ$, $\theta_{J_0}=3.8^\circ$, $\phi_{J_0}=289^\circ$, 
$\phi_\mathrm{c}=305^\circ$, $\alpha_\mathrm{c}=4^\circ$ and $t_\mathrm{c}=700009012.345$\,s GPS time.

\begin{table}
  \caption{
    List of the experiments described in section~\,\ref{sec:skyring}. The sky ring is defined as the ring 
    produced by experiment 1, composed of arcs and gaps.
    \label{table:exp}
  }
  \begin{indented}
  \item[] \begin{tabular}{cccc}
    \br
    experiment  &  injection position  & sky position & starting values \\
                          &  (in the sky ring)     &  parameters & (in the sky ring) \\
    \mr
    1   & arc & free & offset (non true values) \\ 
    2a & arc & fixed &  true position (arc) \\  
    2b & arc & fixed &  gap position (\opencircle)\\  
    3   & gap (\opencircle) & free & offset (non true values) \\    
    \br
  \end{tabular}
  \end{indented}
\end{table}

\begin{figure}  
  \includegraphics[width=0.9\linewidth]{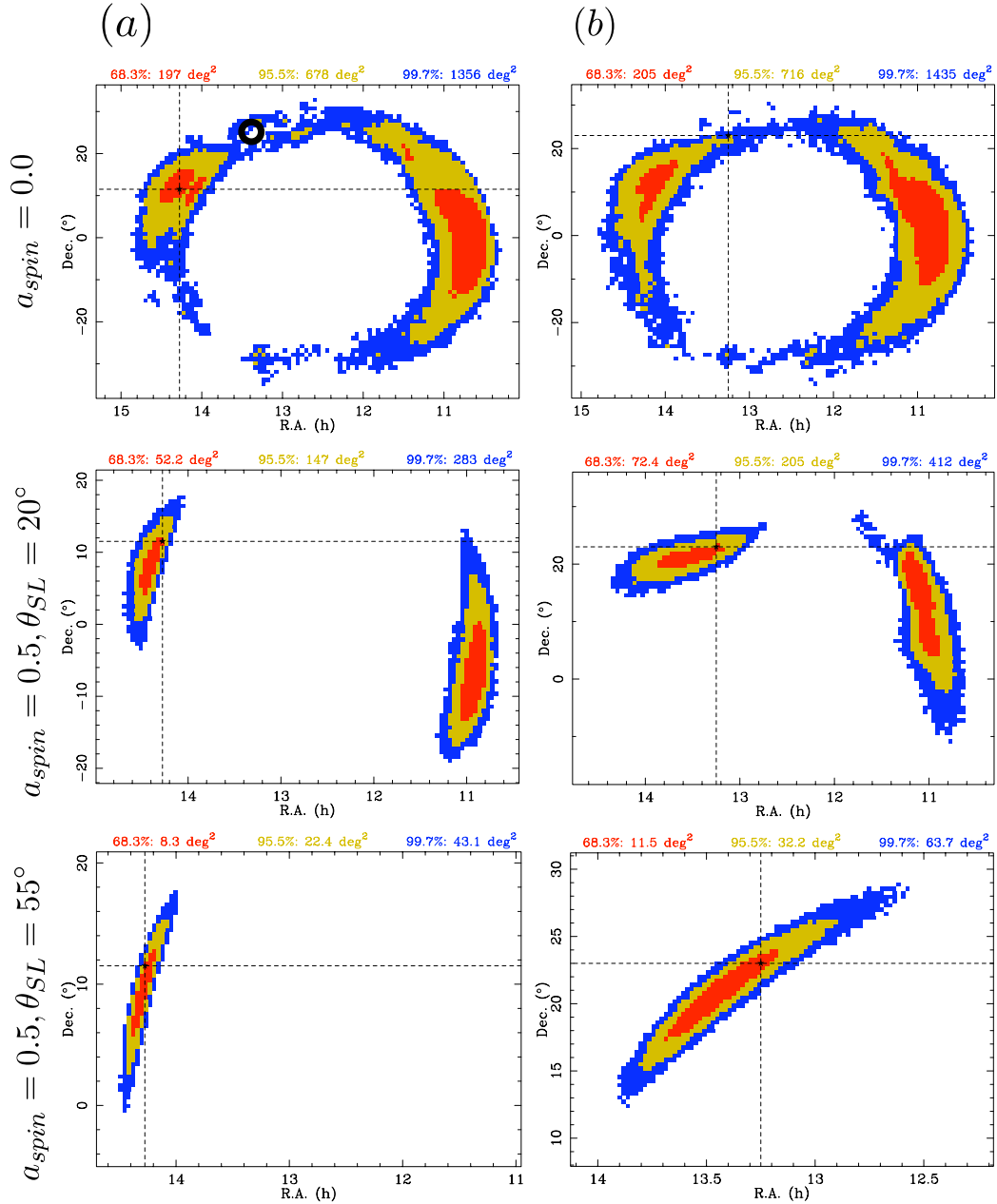}
  \caption{
    Two-dimensional PDFs of the sky position for the MCMC runs as labelled. The colours show the different
    probability intervals (1-$\sigma$, 2-$\sigma$ and 3-$\sigma$ for red, yellow and blue respectively).
    The black dashed lines mark the position 
    in the sky of the injection for each run. Left column {\bf (a)}: results for the reference runs, experiment 1
    (signal injection at R.A.\ = 14.3\,h, Dec = 11.5$^\circ$).  The symbol \opencircle denotes the ``gap'' discussed in the text.
    Right column {\bf (b)}: results for experiment 3: an MCMC run with a signal injection at \opencircle 
    (R.A.\ = 13.25\,h, Dec = 23$^\circ$).
    For the non-spinning case, the PDFs are very similar to those in the original run, whereas they 
    are very different for the spinning cases (notice the difference in the axis ranges).
     \label{fig:sky}
  }
\end{figure}

For a detection with two interferometers, the sky position 
is degenerate; when no spin is present in the source, our PDFs show an incomplete sky ring where the source 
might be --- long arcs separated by empty ``gaps'' --- rather than one tightly constrained ring. When the BH is spinning, and especially when the misalignment angle between orbital angular momentum and spin is significant, 
the sky location constraint shrinks appreciably until much smaller arcs --- or even a single arc --- are left 
(see figure~\,\ref{fig:sky}a).

In order to probe the nature of the gaps in the sky ring, we devised a second experiment 
(experiment 2, see table~\,\ref{table:exp}).  For each run, we 
injected the same signal as before, but now forced the MCMC code to search for the other parameters 
while keeping the sky position fixed to either the true values (experiment 2a), or to the sky position
in the gap that is labelled by the circle in the first panel of figure~\,\ref{fig:sky}a (experiment 2b).
Running the MCMC code in this way provides us with the combination of the free parameters 
that fits the data best, given the constraint in sky position (the {\it conditional} posterior distribution,
conditional on the corresponding sky position). In particular, the code provides us with
the highest likelihood that can be obtained for this sky position. These likelihoods are listed in 
table~\,\ref{table:Lmax}. They show that the maximum likelihood found in the gap is very similar to the 
likelihood of the injection for the non-spinning signal, whereas it is significantly lower in the case of 
the spinning signal. For the non-spinning signal, an unfavourable binary orientation (inclination = 
92.0$^\circ$) and hence a short distance ($d_\mathrm{L}$ = 3.6\,Mpc) are needed to give a good 
match for the given sky position. The modulations of the signal due to the precession of the binary 
orbit prevent this match in the spinning cases.

\begin{table}
  \caption{
    Likelihoods values recovered by the MCMC runs of experiment 2b for the sky ring, described in section~\,\ref{sec:skyring}. 
     \label{table:Lmax}
  }
  \footnotesize
    \begin{tabular}{lcccc}
    \br
                                                                                                                  & $a_\mathrm{spin}=0.0$                             &  $a_\mathrm{spin}=0.5$,  &  $a_\mathrm{spin}=0.5$,  & \\
                                                                                                                  &                                                                        &  $\theta_{SL}=20^\circ$     &  $\theta_{SL}=55^\circ$    & \\
    \mr
    network SNR                                                                                     & 17.0                                                                & 18.3                                       & 18.4                                      & \\
    $\log({\cal L}(x_a|\vec{\lambda}_\mathrm{true}))$                     & 131                                                                  & 154                                        & 208                                      & \\
    $\log({\cal L}_\mathrm{max})$(\opencircle)                                 & 131                                                                  & 125                                        & 152                                       & \\
    Match $M(h(\vec{\lambda}_{\rm true}),h(\vec{\lambda}_{{\cal L}_\mathrm{max}}))$ &  99.5\%   &   89.6\%                                &  82.5\%                               &  (between waveforms injected\\
     (equation~\,\ref{e:M})                                                                       &               
     & & &  in experiment 1 and\\
          &                     &                    &          & those corresponding to ${\cal L}_\mathrm{max}$)\\
                                                                                                                 &  &  &        & \\                                                                                                                                                                                                                  
                                                                                                                  &  &  &                                                                                                                                                                      &(reference parameters,\\
    ${\cal L}_\mathrm{max}$ parameters :                                         & & &                                                                                                                                                                         & injected in experiment 1) \\
    position (R.A. [h], Dec. [$^\circ$])                                                   & 13.25, 23                                                       & 13.25, 23                                & 13.25, 23                             & 14.3, 11.5 \\
    orientation ($\theta_{J_0}$ [$^\circ$], $\phi_{J_0}$ [$^\circ$]) & -65.4, 10.8                                                     & -59.5, 340.4                          & -21.8, 169.5                        & 3.8, 289 \\
    inclination ($\arccos(\mathbf{J} \cdot \mathbf{N})$) [$^\circ$] & 92.0                                                                 & 97.5                                         & 129.7                                   & 128.4 \\ 
    distance [Mpc]                                                                                  & 3.7                                                                   & 11.1                                         & 18.3                                      & 16\\
    $t_\mathrm{c} - $700009010 [s]                                                     & 2.34955                                                       & 2.34959                                   & 2.34960                               & 2.34500 \\
    
    \br
  \end{tabular}
\end{table}

\begin{figure}  
  \includegraphics[angle = 270, width=1.0\linewidth]{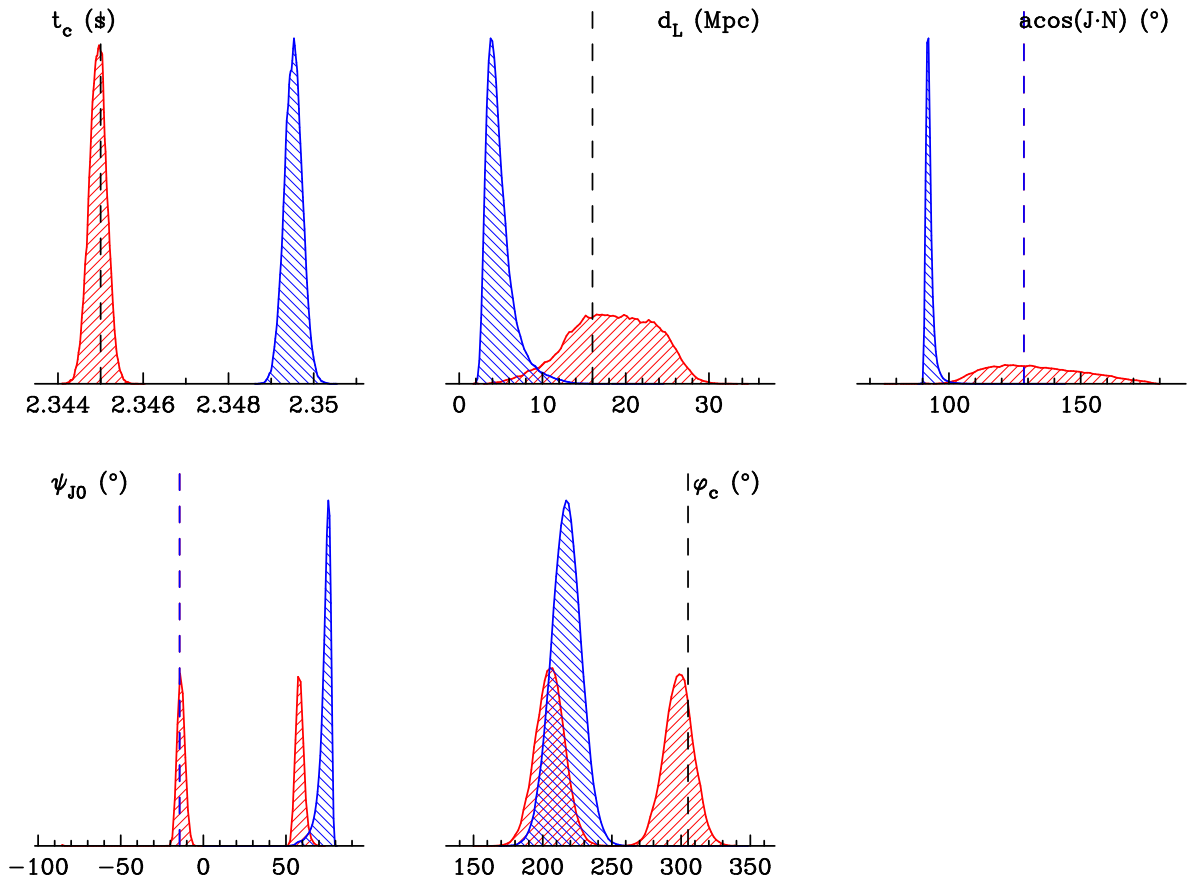}
  \caption{
    One-dimensional PDFs for the runs of experiment 2a (red, hatched upward) 
    and of the runs in experiment 2b (blue, hatched downward), 
    in the non-spinning case. The black dashed lines mark the values of the injected parameters. 
    The level of support is indeed smaller in the second case.
    \label{fig:PDFsky12}
  }
\end{figure}

Thus, for a source with a non-spinning BH, a high likelihood \textit{can} be found in the gaps of the 
sky ring. Therefore the absence of high likelihood values does \emph{not} explain the fact that our 
Markov chains hardly sample this part of the parameter space.
Instead, we find that the PDFs for some of the other parameters (especially the distance $d_\mathrm{L}$ and binary orientation 
$\arccos(\mathbf{J} \cdot \mathbf{N})$) are very narrow for experiment 2b compared to
those for experiment 2a (figure~\,\ref{fig:PDFsky12}). This indicates that 
the overall volume of parameter space in these parameters is very small for sky positions in the gap and this is the reason
that these gaps are not visited frequently by the chains. On a side note, our priors are flat and can not be the cause for this effect.
If our MCMC runs had infinite length, they would
sample this region in parameter space and bridge the gaps in figure~\,\ref{fig:sky}a with a thin line, \textit{i.e.} the points 
``bridging'' the gaps will form a set of very small measure. Even with a perfect sampling, the same behavior is to be expected:
the chains would mainly sample the arcs and rarely the gaps.
Interestingly, the similarity in likelihood values between the true position and the gap for the case of a non-spinning BH
also means that these two signals are virtually indistinguishable, \textit{i.e.} their match (equation~\,\ref{e:M}) is very
high (99.5\% see table~\,\ref{table:Lmax}). This indicates that if the source \textit{were} truly in the gap, Markov chains of this length would not 
recover it, and chains of any length would not have a significant PDF in the gaps.

To illustrate this, we did a third experiment (experiment 3, see table \ref{table:exp}).
We injected a signal at the position in the gap labelled by \opencircle in figure~\,\ref{fig:sky}a. We kept 
the same masses and spin values as before, and set the other parameters to the values yielding the maximum
likelihoods from the second experiment, listed in table~\,\ref{table:Lmax}. 

\begin{figure}  
  \includegraphics[angle = 270, width=1.0\linewidth]{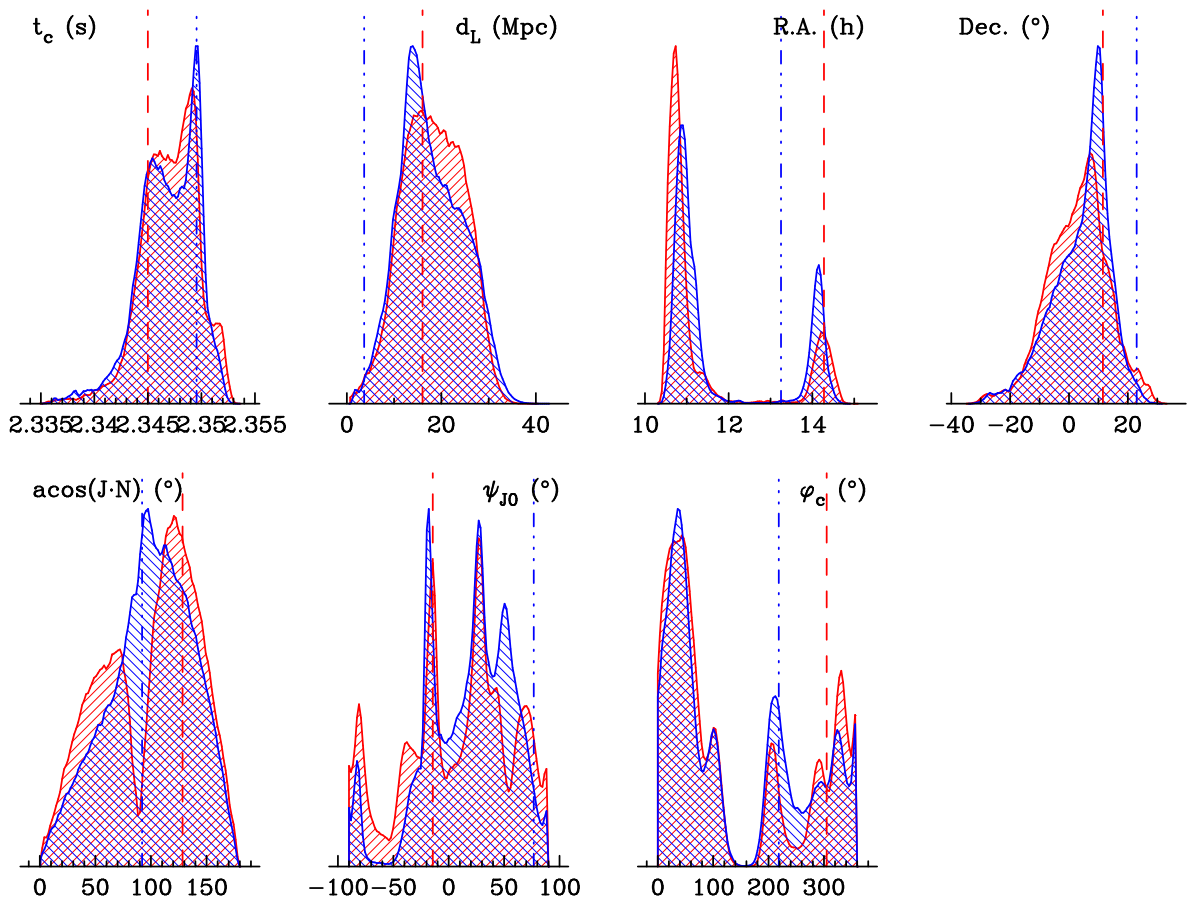}
  \caption{
    One-dimensional PDFs for the reference run in experiment 1 (red, hatched from bottom-left to 
    top-right) and of the runs in experiment 3 (blue, hatched from top-left to bottom-right), 
    in the non-spinning case. The dashed lines mark the values of the injected parameters in experiment 1,
     while the dotted-dashed lines mark the values of the injected parameters in experiment 3.
    The PDFs for all parameters are very similar for the two injections. 
    \label{fig:PDFsky1}
  }
\end{figure}

In the non-spinning case, the sky ring that is recovered is very similar to that of the original run (see figure~\,\ref{fig:sky}b); 
the small difference seen can be explained by the fact that the match between the signal injected in experiment 3 and the signal injected in experiment 1 was 99.5\%, not 100\% (table~\,\ref{table:Lmax}).
This shows that there exist carefully selected combinations of sky position, 
binary orientation and distance which cannot be easily recovered by our analysis. 
However, this reflects a \textit{real} phenomenon; it is very improbable for a source to have the right orientation to 
produce the gravitational-wave signal we injected \textit{and} have its sky location in the gap. And by giving
preference to more likely solutions the parameter estimation routine is in fact doing the right, completely reasonable thing.
The PDFs of the relevant parameters for the comparison of experiments 1 and 3 are plotted in 
figure~\,\ref{fig:PDFsky1}. The PDFs in the other parameters are also very similar in both the original run
and the third experiment run.

When the BH is spinning moderately, the two signals injected in experiment 1 and in experiment 3 
are different (as we can infer from the large 
difference in likelihood in experiment 2b, and the smaller match, see table~\,\ref{table:Lmax}).  The different injections yield significantly different PDFs for the sky position (figure~\,\ref{fig:sky}b) 
as well as for the other parameters, as we checked in our study. 

\subsection{Comparison of MCMC runs with ${1.5}$-pN and ${3.5}$-pN waveforms}
\label{sec:comparison}

So far we have used ${1.5}$-pN waveforms for computational efficiency. However, in this section we used our MCMC implementation to present the first comparison of the accuracy of parameter estimation with the ${1.5}$-pN and ${3.5}$-pN waveform families, as described in section~\,\ref{sec:GW}.  In both cases, we injected BH-NS binary inspiral waveforms with a non-spinning BH into the noise of a single interferometer (H1).  
We analyzed the ${1.5}$-pN injected waveform with an MCMC with ${1.5}$-pN waveform templates, and the ${3.5}$-pN injected waveform with ${3.5}$-pN waveform templates.  We scaled the distance to the source to obtain a signal to noise ratio of 20.0 in both cases, which resulted in a distance 
of $\sim$11.9\,Mpc for the ${1.5}$-pN waveform case, and $\sim$12.2\,Mpc for the ${3.5}$-pN waveform case.
The remaining injection parameters 
were set to $a_\mathrm{spin} = 0.0$, $\Mc=3.0 M_\odot$, $\eta=0.11$ ($M_1=10.0 M_\odot$, $M_2=1.4 M_\odot$), 
R.A.\ = 17.3\,h, Dec = $-5^\circ$, $\theta_{J_0}=-23^\circ$, $\phi_{J_0}=194^\circ$, 
$\phi_\mathrm{c}=352^\circ$ and $t_\mathrm{c}=894377000.500244$\,s for this study. The three spin 
parameters ($a_\mathrm{spin}$, $\theta_{SL}$ and $\alpha_\mathrm{c}$) were fixed, forcing the chains 
to explore a parameter space that was effectively 9-dimensional.

\begin{figure}  
  \includegraphics[angle = 270, width=1.0\linewidth]{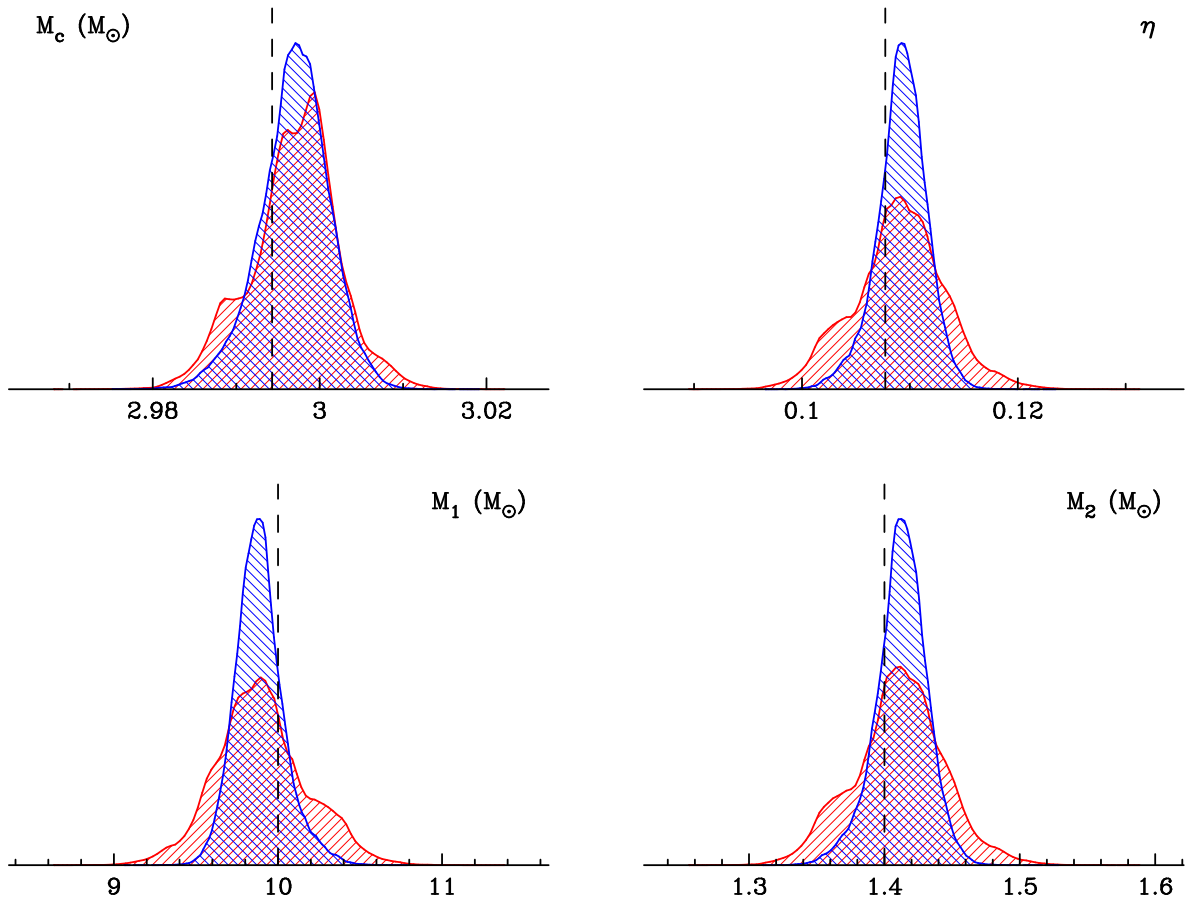}
  \caption{
    PDFs from a run with chains starting from the true values, with the ${1.5}$-pN waveform (red,
    hatched from bottom-left to top-right) and the ${3.5}$-pN waveform (blue, hatched from top-left 
    to bottom-right) from a non spinning source. 
    Only one detector (LIGO at Hanford) is used.
    \label{fig:comppdf}
  }
\end{figure}

Figure~\,\ref{fig:comppdf} compares the probability-density functions (PDFs) of the mass parameters
for the runs with the ${1.5}$-pN and ${3.5}$-pN waveforms for $1.5*10^6$ iterations in both cases. 
It is evident that the estimation of the symmetric mass ratio $\eta$ is more accurate in the ${3.5}$-pN case, 
which results in better constraints on the individual masses.
The 2-$\sigma$ probability ranges for the chirp mass are roughly similar in both cases 
(a factor of 1.2 narrower when the ${3.5}$-pN waveform is used), whereas for $\eta$, and hence 
for the individual masses, the ranges are narrower by a factor of 1.8 when the ${3.5}$-pN waveform is used.
The additional information in the higher-order pN waveforms results in a greater structure 
of the parameter space.  In principle this extra structure allows us to estimate the binary parameters
more accurately. However, a more structured parameter space also affects the sampling efficiency of our 
MCMC code negatively, so that we need more iterations to obtain a well-sampled MCMC run.
In addition, the computation of a single ${3.5}$-pN waveform template takes about 2.4 times longer
than that of a ${1.5}$-pN template. This effect prevents us, for now, from performing this analysis in a full 12-dimensional
or 15-dimensional parameter space in a reasonably short computational time. 

One of the ways to speed up the convergence of the code is to use faster, lower-order pN waveform templates during the \textit{burn-in} phase and to use the additional information provided by the more expensive, higher-order pN waveforms to sample the true mode of the PDF accurately.
In addition, we have plans to use time annealing \cite{2008CQGra..25r4030G}, starting the MCMC runs on only a fraction of the available time-domain data, 
so that only a part of the waveform template needs to be computed for each iteration.
These improvements will speed up the computation of each waveform template and the loss of information will allow the Markov chains 
to move through parameter space more easily until they lock on to the true modes of the PDF. If the amount of information is then gradually increased (by increasing the pN order of the waveforms or the time duration of the data set), the accuracy of parameter estimation will be improved.

\section{Conclusions}
\label{sec:concl}

We have explored the degeneracies in the sky position for a gravitational-wave observation of an inspiral 
of a BH-NS binary with two non-colocated ground-based interferometers. Whereas simple triangulation based on time delays alone would result in a 
homogeneous sky ring, our MCMC runs show (in experiment 1) an incomplete ring in the sky consisting of 
arcs separated by gaps. While the arcs make up most of the circumference of the sky ring for an 
inspiral with a non-spinning BH, these arcs become smaller when spin is present and may be reduced
to a single arc for the case of moderate spin and a sufficient misalignment between the BH spin and the orbital angular momentum. We demonstrated
that in the spinning case, the maximum likelihood values are in fact lower in the gaps than in the arcs.  
In the non-spinning case the likelihood in the gaps can actually be as high as in the arcs; however the
gaps can be explained by a smaller volume of support in the other extrinsic parameters, such as the binary orientation and distance (experiment 2). 
It is then less likely for the chains to sample this smaller volume in parameter space, resulting in gaps in the two-dimensional sky probability density function.
In the non-spinning case, if a source is located in a gap, the 
posterior PDF still has a gap at the true source sky location
(experiment 3).

The subject of estimating the position of a gravitational-wave source on the sky has been explored by many researchers,
in the context of binary-inspiral and burst signals; the assumed baseline for these studies corresponds
to three detectors located at the three LIGO-Virgo observatory positions, and operating at their design sensitivities. For 
detectable burst sources (SNR$>$5), arrival-time techniques should allow a precision of about $1^\circ$ for the source
direction~\cite{2006PhRvD..74h2004C}. A method that takes into account burst-signal arrival time and amplitude, plus arrival-time
uncertainties equivalent to what is observed with real LIGO and Virgo data, gives source uncertainties of a few
degrees~\cite{2008PhRvD..78l2003M}. Coherent techniques also exist for burst detection and sky-location determination, 
but by their own admission these methods are computationally costly~\cite{2008arXiv0809.2809S}. Many techniques have also been 
developed for sky localization for binary-inspiral events. It has been shown, using time and mass parameter
estimates, that the LIGO-Virgo network could localize the position of a binary-neutron-star inspiral to an accuracy of 
$4^\circ$ if it were located in M87 (16 Mpc), or $2^\circ$ for a source in NGC 6744 (10 Mpc)~\cite{2008CQGra..25d5001B}; coherent methods,
which also depend on estimating the time and mass parameters, give similar accuracies~\cite{2009arXiv0901.4936A}.

MCMC parameter-estimation methods, like those used in this study, are capable of estimating the sky parameters,
along with all of the other signal parameters; this is one reason why MCMC methods are computationally intensive. Coherent MCMC
methods applied to signals observed by the LIGO-Virgo network will be able to resolve the sky location to ~$2^\circ$ for signals with
an SNR of 15~\cite{2007CQGra..24..607R}. When the compact objects have spin, and the search templates account for this parameter, sky
localization becomes relatively more accurate for higher values of spin~\cite{2008ApJ...688L..61V}.
It is important to remember that, in principle, the MCMC results show the best constraints one could hope to place on signal parameters (including the sky location) 
by displaying the true posterior PDFs. While comparisons with other particular sky-localization results may be cumbersome because different waveforms 
were assumed at different SNRs in different detectors, MCMC produced posterior PDFs display the statistically correct and most precise localization.
Bayesian methods achieve better parameter-estimation accuracy when the template model describes the functional form 
of the actual signal more accurately. When a gravitational-wave detection occurs, it is likely 
that all possible sky localization algorithms will be used, and the methods should be considered to be complementary.

In section~\ref{sec:comparison}, we compared MCMC results on software injections using waveform families 
of ${1.5}$-pN and ${3.5}$-pN order for both the injections and the MCMC parameter estimation. We have shown that the higher-order 
templates have the potential for more accurate parameter estimation, but that sampling the parameter space with these templates 
is more computationally difficult. However, a number of upcoming improvements should improve the sampling efficiency of our MCMC runs.

The analysis presented here is the second step of a more detailed  
study that we are currently carrying out, exploring a much larger parameter space,  
developing techniques to reduce the computational cost of these simulations, and  
testing the methods with actual LIGO data. We are in the process of updating our MCMC code to include 
the spin of the second binary member, increasing the dimensionality of the parameter space from 12 to 15. 
Finally, we intend to further develop our Bayesian approach into a standard  
tool that can be included in the follow-up analysis pipeline used for the processing of the  
`science data' collected by ground-based laser interferometers.

\ack

The authors would like to thank Gareth Jones for his help in the understanding of the LAL parameter system.
This work is partially supported by the National Science Foundation grant NSF-0838740,
a NSF Gravitational Physics grant (PHY-0653321) to VK; NSF Gravitational Physics
grant (PHY-0553422) to NC and the Max-Planck-Society (CR).
Computations were performed on the Fugu computer cluster funded by NSF MRI grant PHY-0619274 to VK.

\section*{References}
\bibliography{mcmc-NRDA}{}
\bibliographystyle{jphysicsB}

\end{document}